\documentclass[conference,10pt]{IEEEtran}

\usepackage[top=0.75in, bottom=1.03in, left=0.625in, right=0.625in]{geometry}
\usepackage{epsfig,rotating,setspace,latexsym,amsmath,epsf,amssymb,amsfonts,bm,theorem,subfigure,epstopdf}

\usepackage{cite}
\usepackage{bbm}
\usepackage{enumerate} 
\usepackage{layout}

\usepackage[ruled,vlined,linesnumbered]{algorithm2e}
\usepackage{subcaption} 
\usepackage{multirow}
\usepackage{listings}
\usepackage{makecell}
\usepackage{graphics}
\usepackage{graphicx}
\usepackage{adjustbox}
\usepackage{array}

\newcolumntype{?}{!{\vrule width 1pt}}

\usepackage{color,booktabs,multirow}

\IEEEoverridecommandlockouts
\allowdisplaybreaks
\bstctlcite{IEEEexample:BSTcontrol}
\begin{document}


    \title{Deep Reinforcement Learning for Multi-flow Routing in Heterogeneous Wireless Networks}
	\author{\IEEEauthorblockN{Brian Kim, Justin H. Kong, Terrence J. Moore, and Fikadu T. Dagefu}
 \IEEEauthorblockA{U.S. Army DEVCOM Army Research Laboratory, Adelphi, MD, 20783, USA}
	\thanks{A preliminary version of the material in this paper has partially appeared in Proceedings of IEEE Military Communications Conference (MILCOM) 2025~\cite{kim2025single}.}

 }

\maketitle


\begin{abstract}

Due to the rapid growth of heterogeneous wireless networks (HWNs), where devices with diverse communication technologies coexist, there is increasing demand for efficient and adaptive multi-hop routing with multiple data flows. Traditional routing methods, designed for homogeneous environments, fail to address the complexity introduced by links consisting of multiple technologies, frequency-dependent fading, and dynamic topology changes. In this paper, we propose a deep reinforcement learning (DRL)-based routing framework using deep Q-networks (DQN) to establish routes between multiple source-destination pairs in HWNs by enabling each node to jointly select a communication technology, a subband, and a next hop relay that maximizes the rate of the route. Our approach incorporates channel and interference-aware neighbor selection approaches to improve decision-making beyond conventional distance-based heuristics. We further evaluate the robustness and generalizability of the proposed method under varying network dynamics, including node mobility, changes in node density, and the number of data flows. Simulation results demonstrate that our DRL-based routing framework significantly enhances scalability, adaptability, and end-to-end throughput in complex HWN scenarios.

\end{abstract}


\section{Introduction}

The rapid expansion of wireless technologies has led to increasingly diverse wireless network environments, where a wide range of devices with different capabilities, protocols, and operating standards coexist. These heterogeneous wireless networks (HWNs) integrate devices that use various communication technologies, utilizing different parts of the spectrum, offering distinct trade-offs in terms of data rate, range, directionality, and varying characteristics in line-of-sight (LOS) and non-line-of-sight (NLOS) conditions. In such networks, multi-hop communication is often used to extend the communications coverage and to ensure connectivity in environments with a high level of channel dynamics

Conventional routing of multi-hop communications typically assumes homogeneous devices and uniform wireless links. A well-known traditional method for routing in such networks is the open shortest path first protocol using Dijkstra's algorithm \cite{dijkstra1959note} to find the shortest path to the destination. However, this approach, as well as other conventional approaches including Destination-Sequenced Distance Vector routing (DSDV) \cite{perkins1994highly} and Ad hoc On-Demand Distance Vector (AODV) \cite{perkins1999ad}, lack adaptability to heterogeneous networks, especially when each node is equipped with diverse wireless communications technologies. Recently, a node using multiple technologies has been introduced as cross-technology communication (CTC) \cite{chen2019survey} which significantly expands the routing space, allowing for more flexible and efficient path selection. However, leveraging such nodes introduces new challenges in routing and dynamic technology selection since route decisions in such environments must consider heterogeneity not only at the device level but also at the link level, accounting for differences in bandwidth, interference, and channel effects. Also, the overhead of collecting all necessary information for routing in advance is substantial as the network size grows \cite{zhou2003reactive}, motivating the need to develop a scalable and robust routing algorithm for multi-hop HWNs.

To address the limitations of conventional routing approaches, decentralized algorithms have been proposed for multi-hop wireless networks to reduce the overhead at each node while accommodating diverse network objectives and dynamic environments \cite{kandelusy2020cognitive,senanayake2018decentralized}. In these approaches, each node independently selects the next hop by balancing multiple decision factors, making the routing process a sequential decision-making problem. Here, each decision solely depends on the current state of the network, which includes only the information from neighboring relay nodes, making the algorithm decentralized. This sequential and state-dependent structure naturally aligns with the framework of a Markov Decision Process (MDP) \cite{bellman1957markovian}, making reinforcement learning (RL) a promising solution.

Q-learning, a model-free value-based RL algorithm, has been applied to determine optimal routes in multi-hop wireless networks \cite{chang2004mobilized,arafat2021q}. In this approach, each node maintains a table of Q-values representing the expected utility of selecting a particular neighbor as the next hop, with the Q-values being iteratively updated based on the rewards received from taking actions. This method has been extended to HWNs, where multiple communication technologies are employed to enhance rate, reduce latency, and improve covertness \cite{kong2024decentralized,kim2024failure}. However, Q-learning-based routing approaches in HWNs face significant scalability and generalization challenges. Specifically, each node must store a large Q-value table tailored to a particular network topology, becoming inefficient and impractical when network conditions change dynamically.

Thus, a deep RL (DRL) approach has been introduced to handle routing decisions in a more robust and scalable manner for multi-hop wireless networks. DRL leverages deep neural networks (DNNs) to approximate decision policies, allowing routing strategies to dynamically adapt to changing network conditions such as node mobility and topological variations, without being confined to a fixed network structure. In addition, DRL-based methods can incorporate diverse quality-of-service (QoS) metrics, such as bandwidth, latency, and interference, into the routing process \cite{zhao2020deep, rathore2019deep}. Among various DRL techniques, the deep Q-learning (DQL) is particularly notable, which employs DNNs as the Q-value function by enabling nodes to learn optimal policies. This reduces the need to store an extensive table with Q-values at each node and allows the model to generalize across different network topologies, making routing more efficient even in dynamic and large-scale multi-hop networks. Despite these advantages, existing DRL-based routing approaches often rely on simplistic, distance-based channel models, limiting their applicability in real-world scenarios where fading effects arise due to the specific environmental geometry and material properties. Furthermore, as HWNs increasingly employ multiple communication technologies to improve coverage and data rates, it remains an open question how DRL can fully leverage these heterogeneous capabilities for efficient and adaptive routing. 


To this end, we propose a DRL-based routing framework for multi-flow HWNs, extending our prior work \cite{kim2025single} which focused on a single-flow scenario. In the proposed framework, each node aims to maximize the overall sum rate of multiple data flows through a two-stage process: 1) neighbor node set selection, 2) route decision and resource allocation. Each node first selects neighbor nodes as a set of potential next hops where we propose several neighbor node set selection strategies tailored for heterogeneous environments, aiming not only to reduce communication overhead associated with information exchange but also to address the limitations of relying solely on physical proximity. In particular, selecting the nearest nodes may not yield optimal performance in scenarios where channel quality is impacted by frequency-dependent local fading effects. Then, in the route decision and resource allocation process, each node jointly selects both a communication technology (along with its available subbands) and a next-hop relay from the set of neighboring nodes using the DRL agent. 

During routing, our approach takes not only interference from other flows but also interference within the flow, while additionally incorporating the distinct channel characteristics inherent to different communication technologies, which arise from their varying operating frequencies. While DRL-based methods have shown the ability to adapt to certain dynamic environments, to the best of our knowledge, their ability to effectively handle node mobility and network change in HWNs remains underexplored. Understanding how mobility impacts routing decisions is crucial for maintaining connectivity and optimizing resource utilization in dynamic scenarios. This motivates a thorough investigation into the robustness of the pretrained DRL-based routing to mobile nodes. Furthermore, we aim to investigate the generalizability of the pretrained DRL-based routing where the density of nodes in the topology and the number of data flows change from what DRL has observed during the training. The key contributions of the paper are

\begin{itemize}
    \item \textbf{DRL-based routing for heterogeneous wireless networks with multiple flows:}
We propose a novel DRL-based routing framework that enables each node to jointly select the communication technology, subband, and next-hop relay, aiming to maximize the sum rate of all flows while accounting for interference, frequency-dependent fading, and link heterogeneity. 
    \item \textbf{Frequency-aware and interference-aware neighbor selection strategies:}
We introduce and evaluate various neighbor selection methods that go beyond proximity-based heuristics, incorporating local channel conditions and interference patterns, which differ across technologies and frequencies in HWNs.
    \item \textbf{Mobility-aware generalizable routing using DRL:}
We investigate the performance and generalization capability of a pretrained DRL agent under node mobility, change in density of nodes in the topology, and in network density, providing insights into their adaptability and robustness in dynamic network environments.
\end{itemize}

The rest of the paper is organized as follows. Section \ref{sec:related work} presents a brief background of the related works. Section \ref{sec:system model and problem formulation} describes the system model and formulation of the problem that we aim to solve. Section \ref{sec:DRL} describes our neighbor selection approaches, state and action spaces, and the architecture of the DRL agent. Section \ref{sec:simulations} shows the performance of the DRL agent considering different environments and the robustness against node mobility and topology dynamics. Section \ref{sec:Conclusion} concludes the paper.

\section{Related Works} \label{sec:related work}
Most traditional routing algorithms for wireless networks, as discussed in \cite{gafni2003distributed, chen1999distributed, dube1997signal}, have primarily relied on heuristic approaches. In contrast, recent advances demonstrate that RL can significantly enhance routing performance by enabling learning-based decisions that adjust to network conditions, leveraging the inherently sequential nature of the routing process. Previous works have considered Q-learning \cite{watkins1992q}, a type of RL, to find the best route by using the saved Q-values that are updated using the reward function composed of targeted network features such as residual energy \cite{yun2021q}, rate \cite{al2014reinforcement,al2016routing}, communication cost \cite{serhani2016qlar}, and security \cite{kong2024decentralized,kim2025reinforcement}. Adapting to the changed environment using Q-learning has been studied in \cite{arafat2021q,kim2024failure} by adaptively adjusting Q-learning parameters (e.g., learning rate and discount factor) during the retraining phase, which has shown improved results in terms of convergence speed and end-to-end delay. Further, routing has been considered in HWNs where each node is equipped with multiple wireless communication technologies operating at different frequency bands with unique channel characteristics  \cite{kong2024decentralized,kim2025reinforcement}. However, due to the inherent nature of Q-learning, these works need to store Q-values for a specific topology and source-destination pair causing scalability issues when applied to larger networks and multiple flow scenarios.

DRL has also been proposed, leveraging DNNs, to enhance the scalability and robustness in wireless networks. For instance, \cite{cui2021scalable} investigates simultaneous routing and spectrum access in multi-flow wireless networks, where each node employs a DRL agent to jointly determine routing and spectrum allocation for a single communication technology. 
The objective is to minimize the end-to-end signal-to-interference-plus-noise ratio (SINR) by considering physical-layer characteristics of a homogeneous network such as interference, distance, and angle. This work employs deep Q-networks (DQNs) \cite{mnih2015human}, a prominent DQL method that uses DNNs to approximate Q-values enabling nodes to learn optimal policies through experience replay and target networks. Another DRL approach has been considered in \cite{paul2023multi} where graph neural networks (GNNs) have been used to establish multiple routes for wireless interference networks. Here, the goal is to maximize the network utility which is closely related to the sum rate of multiple flows where the proposed approach consists of two parts, a centralized part and a GNN-based decentralized part. Further, \cite{zhang2024decentralized} explores joint optimization of routing and transmit power using DRL for multi-flow networks, incorporating both rate and path connection probability as performance metrics. Collectively, these works demonstrate that DRL can adapt effectively to dynamic topologies when trained across diverse network topologies under distance-based channel models.
However, to the best of our knowledge, using DRL for routing in HWNs has not been explored and needs investigation due to the increase in complexity from multiple communication technologies at each node where each communication technology has its own characteristics.
Moreover, there is limited work focusing on the robustness of the DRL technique in dynamic network conditions (e.g., node mobility) and the generalizability of the DRL technique to unseen network conditions.


\section{System Model and Problem Formulation} \label{sec:system model and problem formulation}
\label{sec:SystemModel}

\begin{figure*}[t]
    \centering  \includegraphics[scale=0.2]{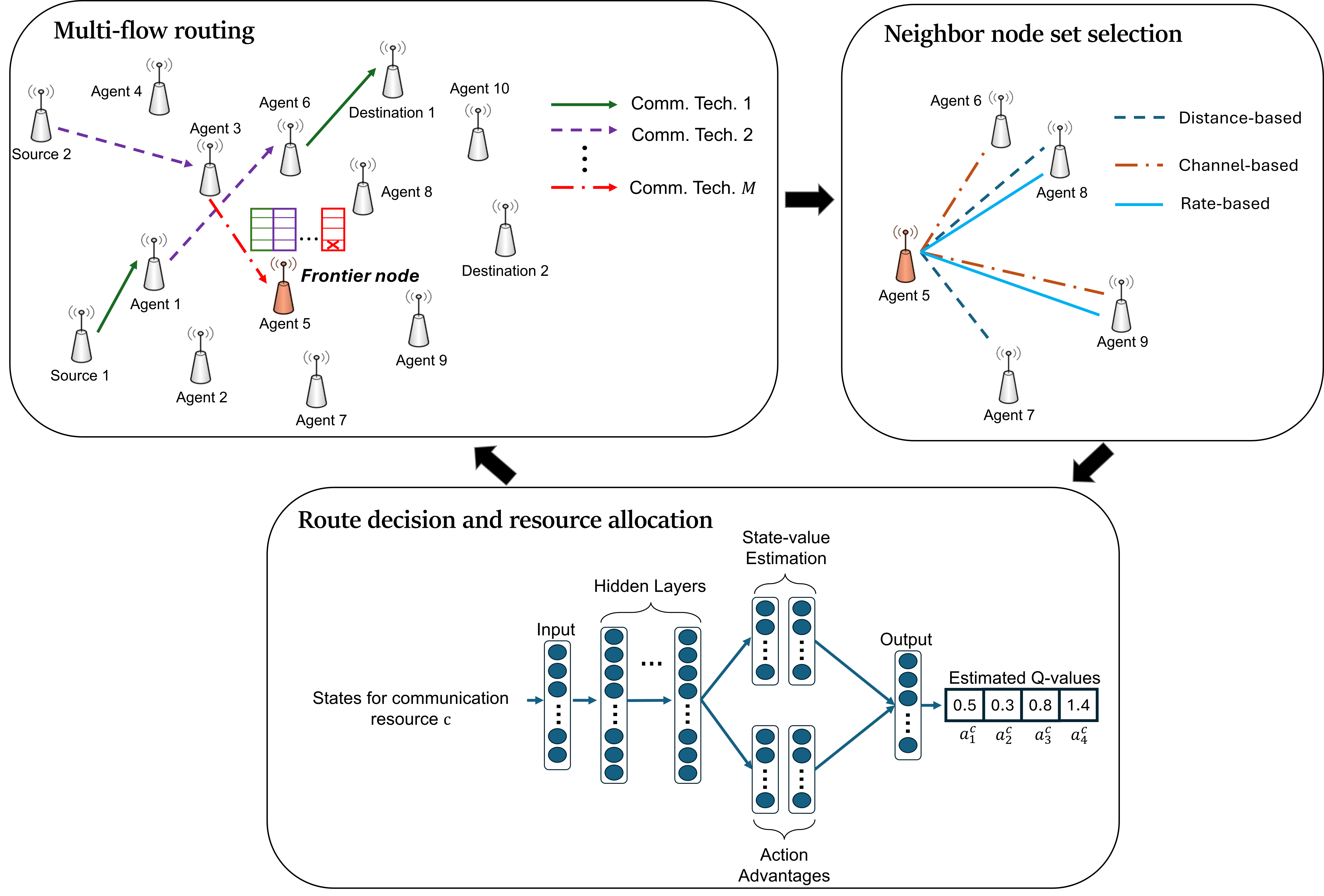}
    \caption{Overall framework of DRL-based HWN routing for multi-flow. The frontier node decide the next hop and communication resource through a two-stage process: 1) neighbor node set selection, 2) route decision and resource allocation.}
    \label{fig:framework}
\end{figure*}

\subsection{Network Model} \label{sec: network model}
We consider an HWN that supports $F$ data flows, consisting of $F$ number of distinct source-destination pairs, and $E$ relay nodes, a subset of which will constitute routes between multiple source-destination pairs. Note that all sources and destinations are not used as relay nodes for other data flows and are only dedicated to their own data flow. All nodes, including sources and destinations in the network, are equipped with $M$ different communication technologies, as seen in Fig. \ref{fig:framework}, and each communication technology $m$ has $B^{(m)}$ available subbands. Here, each communication technology uses a unique frequency band and has its own channel characteristics.

We first define $\mathcal{G}^{(f)}$ as the set of all possible routes that can be formed for the $f$-th flow. Then, a possible route for the $f$-th flow is defined as 
\begin{equation}
G^{(f)} = \{e^{(f)}_0,e^{(f)}_1, e^{(f)}_2, \cdots,e^{(f)}_{N_{G^{(f)}}},e^{(f)}_{N_{G^{(f)}}+1} \} \in \mathcal{G}^{(f)},
\end{equation}
where $e^{(f)}_{i} \in \mathcal{E}$ is the $i$-th relay node in the flow $f$ among the set of relay nodes $\mathcal{E}$ with $|\mathcal{E}| = E$. $N_{G^{(f)}}$, $e^{(f)}_0$, and $e^{(f)}_{N_{G^{(f)}}+1}$ are the number of relay nodes used for flow $f$, the source node, and the destination node, respectively.

Furthermore, a possible set of communication technologies and subbands used for the $f$-th route from the all possible set $\mathcal{C}^{(f)}$ is defined as 
\begin{equation}
    C^{(f)} = \{c^{(f)}_{1},c^{(f)}_{2}, \cdots,c^{(f)}_{N_{G^{(f)}}+1} \}  \in \mathcal{C}^{(f)},
\end{equation}
where $c^{(f)}_{i}\in \mathcal{C}$ is the communication technology and corresponding subband used at the $i$-th hop and $\mathcal{C}$ is the set of all communication resources (combination of all available communication technologies and subbands), defined as
\begin{equation}
    \mathcal{C} = \{c_{(1,1)},c_{(1,2)}, \cdots, c_{(M,B^{(M)}-1)},c_{(M,B^{(M)})} \},
\end{equation}
with $|\mathcal{C}| = \sum_{i=1}^{M} B^{(i)}$. Here, $c_{(i,j)}$ represents the communication technology $i$ using subband $j$, and we assume that there exists no interference between different communication resources.

When choosing the communication resource at each node, we only allow one communication resource for transmission at the node and assume that a node cannot simultaneously transmit and receive. Also, we restrict each flow to relay through a node at most once to avoid cycles in a route. Mathematically, for the node $e^{(f)}_i$,
\begin{align}
     &c^{(f)}_{i} \neq c^{(f)}_{i+1},\\
     &e^{(f)}_{i} \ne e^{(f)}_{j}, \;\;\;\; \forall j \ne i.
\end{align}

In the meantime, we allow each relay node to serve multiple routes simultaneously as long as all transmission and reception signals use different communication resources. 
If $e^{(f_1)}_i = e^{(f_2)}_j$ for some $f_1, f_2 \in \mathcal{F}$ and some $i$ and $j$, then
\begin{equation}
c^{(f_1)}_{i} \neq c^{(f_1)}_{i+1} \neq  c^{(f_2)}_{j} \neq c^{(f_2)}_{j+1}.
\end{equation}

\subsection{Performance Metric}

When the transmitter node $e^{(f)}_i \in G^{(f)}$ in the route $G^{(f)}$ selects a resource $c^{(f)}_{i}$ from $\mathcal{C}$ and sends data symbol $x_{e_{i}}^{(f)} \sim \mathbb{C}(0,1)$, the received signal at the receiver $e^{(f)}_{i+1}$ is
\begin{align}\label{eq:Received_Receiver}
    y_{e^{(f)}_{i+1}}^{c^{(f)}_{i}} = &\sqrt{P} h_{e^{(f)}_i,e^{(f)}_{i+1}}^{c^{(f)}_{i}} x_{e_{i}}^{(f)} 
    +\underbrace{\sum_{ \substack{c^{(q)}_{k}=c^{(f)}_{i} \\ q \neq f}}\sqrt{P} h_{e^{(q)}_k,e^{(f)}_{i+1}}^{c^{(q)}_{k}} x_{e_{k}}^{(q)}}_{\text{IFI}} \nonumber \\
    &+\underbrace{\sum_{\substack{c^{(f)}_{k}=c^{(f)}_{i} \\ k \ne i}}\sqrt{P} h_{e^{(f)}_k,e^{(f)}_{i+1}}^{c^{(f)}_{k}} x_{e_{k}}^{(f)}}_{\text{IHI}}  + n_{e^{(f)}_{i+1}}^{c^{(f)}_{i}},
\end{align}
where $P$ is the transmit power at the transmitter, $h_{e^{(f)}_i,e^{(f)}_{i+1}}$ denote the channel from transmitter $e^{(f)}_i$ to receiver $e^{(f)}_{i+1}$ 
for resource $c^{(f)}_i$. Further, $n_{e^{(f)}_{i+1}}^{c^{(f)}_{i}} \sim \mathcal{CN}(0,\Omega^{c^{(f)}_{i}}N_0)$ is AWGN at the receiver 
where $\Omega^{c^{(f)}_{i}}$ is the bandwidth for communication resource $c^{(f)}_{i}$ 
and $N_0$ is the noise power spectral density.
Then, the SINR at the receiver $e^{(f)}_{i+1}$ for resource $c^{(f)}_{i}$ is defined as 
\begin{align}\label{eq:SINR}
&\text{SINR}^{c^{(f)}_{i}}_{e^{(f)}_{i+1}} = \nonumber \\
&\frac{P | h_{e^{(f)}_{i},e^{(f)}_{i+1}}^{c^{(f)}_{i}} |^2}{\sum\limits_{ \substack{c^{(q)}_{k}=c^{(f)}_{i}\\ q \neq f}}P |h_{e^{(q)}_k,e^{(f)}_{i+1}}^{c^{(f)}_{k}}|^2 + \sum\limits_{\substack{c^{(f)}_{k}=c^{(f)}_{i}\\ k \ne i}}P |h_{e^{(f)}_k,e^{(f)}_{i+1}}^{c^{(f)}_{k}}|^2 + \Omega^{c^{(f)}_{i}}N_0},
\end{align}
and the rate of a single-hop link between nodes $e^{(f)}_i$ and $e^{(f)}_{i+1}$ is given by
\begin{align} \label{eq:Rate_link}
	R^{c^{(f)}_{i}}_{e^{(f)}_{i},e^{(f)}_{i+1}} = \Omega^{c^{(f)}_{i}} \log_2\big(1+\text{SINR}&^{c^{(f)}_{i}}_{e^{(f)}_{i+1}}\big).
\end{align}
Since the lowest rate of the single-hop links in the route determines the overall rate, the end-to-end rate of route $G^{(f)}$ is defined as
\begin{align} \label{eq:Rate_route}
	&R(G^{(f)}, C^{(f)}) =  {\min\limits_{ \substack{e^{(f)}_i \in G^{(f)}\\ c_i^{(f)} \in C^{(f)}}}} R^{c^{(f)}_{i}}_{e^{(f)}_{i},e^{(f)}_{i+1}},
\end{align}
and the sum rate of all routes is 
\begin{equation}
    R_{tot} = \sum\limits_{f = 1}^{F} R(G^{(f)},C^{(f)}) .
\end{equation}

\subsection{Optimization Problem Formulation}
In this paper, we aim to maximize the sum rate of $F$ flows by carefully determining the routes and the communication resources used for the transmissions. Mathematically, the optimization is formulated as  
\begin{align} \label{eq:objective function}
    \underset{ \substack{ {G}^{(f)} \in \mathcal{G}^{(f)}, \\ {C}^{(f)} \in \mathcal{C}^{(f)}, \forall f \in \mathcal{F}}}{\max} & \;\; \sum\limits_{f = 1}^{F} R(G^{(f)},C^{(f)}) \\ 
	 \text{s.t.} \;\;\;\;\;\;\;&  e^{(f)}_{i} \ne e^{(f)}_{j}, \;\;\;\; \forall j \ne i, \;\; \forall f \in \mathcal{F} \nonumber \\
      c^{(f)}_{i}& \neq c^{(f)}_{i+1} \neq  c^{(f')}_{j} \neq c^{(f')}_{j+1} \;\;   \forall j \ne i, \;\; \forall f, f' \in \mathcal{F}. \nonumber
\end{align} 
Note that the optimization problem (\ref{eq:objective function}) is an integer programming problem, which is an NP-hard problem with a large number of discrete optimization variables involved which grows as the number of nodes and communication resources involved in the wireless network increases. Further, the problem is highly complex due to the interdependency between flows and hops caused by interference as expressed in~(\ref{eq:Received_Receiver}). Specifically, if two hops from the same flow or from other flows use the same communication resource, then the transmitted signal of a hop interferes with the other hop, making spectrum allocation and communication technology selection crucial for the routing problem. Also, using a node for a flow leads to a decrease in potential decision space for the other flows due to the limitation of each node participating in multiple flows. This interdependence among routing decisions, spectrum allocation, and interference management significantly increases the computational complexity of the problem, necessitating the use of advanced heuristic techniques to obtain feasible and near-optimal solutions.

To tackle this difficulty, we model the problem as an MDP, since the routing problem involves sequential decision-making in a dynamic environment. In this paper, we assume that each node observes local information from neighbor nodes without the help of a centralized server and decides the action in a distributed manner. This makes it a partially observable Markov decision process (POMDP), which motivates the use of multi-agent RL (MARL) where each node acts as an agent. Furthermore, unique characteristics of the wireless environment due to the wireless medium, such as dynamic channel conditions and interference, motivate the need to come up with a DRL agent robust to highly varying wireless networks.

\begin{table}
\centering
\caption{Notations used in this work.\label{tab:notations}}
\begin{adjustbox}{width=0.93\columnwidth, center}
{%
\begin{tabular}{@{}cp{50mm}@{}}
\toprule
\textbf{Notation}     &\textbf{Description}   \\
\midrule
$F$ & Num. of data flows  \\ \cline{1-2}
$E$ & Num. of relay nodes     \\ \cline{1-2}
$B^{(m)}$ & \multicolumn{1}{l}{\begin{tabular}[c]{@{}l@{}} Num. of subbands for comm. tech. $m$\\   \end{tabular}}   \\ \cline{1-2}
$\Omega^{(m)}$ & \multicolumn{1}{l}{\begin{tabular}[c]{@{}l@{}} Bandwidth of a subband for comm. tech. $m$  \end{tabular}}  \\ \cline{1-2}
$\tau_{m}$ & \multicolumn{1}{l}{\begin{tabular}[c]{@{}l@{}} Center frequency of comm. tech. $m$  \end{tabular}}  \\ \cline{1-2}
$\mathcal{G}^{(f)}$ & \multicolumn{1}{l}{\begin{tabular}[c]{@{}l@{}}  Set of all possible routes for flow $f$  \end{tabular}}   \\ \cline{1-2}
$G^{(f)}$ & One possible route for flow $f$ \\ \cline{1-2}
$e^{(f)}_i$ & $i$-th relay node in flow $f$  \\ \cline{1-2}
$\mathcal{E}$ & Set of all relay nodes  \\ \cline{1-2}
$N_{G^{(f)}}$ & \multicolumn{1}{l}{\begin{tabular}[c]{@{}l@{}} Num. of relay nodes used for  flow $f$ \end{tabular}}   \\ \cline{1-2}
$C^{(f)}$ & \multicolumn{1}{l}{\begin{tabular}[c]{@{}l@{}} One possible comm. resources used for flow $f$ \end{tabular}}  \\ \cline{1-2}
$c_{i}^{(f)}$ & \multicolumn{1}{l}{\begin{tabular}[c]{@{}l@{}} Comm. resource used at the $i$-th hop\end{tabular}}\\ \cline{1-2}
$c_{(i,j)}$ & \multicolumn{1}{l}{\begin{tabular}[c]{@{}l@{}} Comm. tech. $i$ using subband $j$\end{tabular}}    \\ \cline{1-2}
$x^{(f)}_{e_i}$ & Data sent at node $e_i^{(f)}$  \\ \cline{1-2}
$y^{c^{(f)}_{i}}_{e_i^{(f)}}$ & Received signal at node $e_{i}^{(f)}$  \\ \cline{1-2}
$P$ & Transmit power  \\ \cline{1-2}
$h_{\alpha,\beta}^{c_{i}^{(f)}}$ & \multicolumn{1}{l}{\begin{tabular}[c]{@{}l@{}}Channel from Tx $\alpha$ to Rx $\beta$ for resource $c^{(f)}_ i$ \end{tabular}}     \\ \cline{1-2}
$N_0$ & Noise power  \\ \cline{1-2}
$\text{SINR}^{c^{(f)}_{i}}_{e^{(f)}_{i}}$ & \multicolumn{1}{l}{\begin{tabular}[c]{@{}l@{}}SINR at the receiver $e^{(f)}_{i}$ for resource $c^{(f)}_{i}$\end{tabular}}    \\ \cline{1-2}
$R^{c^{(f)}_{i}}_{e^{(f)}_{i},e^{(f)}_{i+1}}$ &  \multicolumn{1}{l}{\begin{tabular}[c]{@{}l@{}}Rate of a single-hop link between nodes\\ $e^{(f)}_i$ and $e^{(f)}_{i+1}$ \end{tabular}}    \\ \cline{1-2}
$R(G^{(f)}, C^{(f)})$ & End-to-end rate of route $G^{(f)}$   \\ \cline{1-2}
$R_{tot}$ & Sum rate of all routes \\ \cline{1-2}
$\mathcal{A}_e^{(c)}$ & \multicolumn{1}{l}{\begin{tabular}[c]{@{}l@{}}Set of all possible actions at frontier\\  node $e$ using comm. resource $c$ \end{tabular}}   \\ \cline{1-2}
$a_{\eta_i}^{(c)}$ & \multicolumn{1}{l}{\begin{tabular}[c]{@{}l@{}}Action of transmitting data to $i$-th neighbor\\   node $\eta_i$ using comm. resource $c$\end{tabular}}  \\ \cline{1-2}
$\eta_i$ & $i$-th neighbor node \\ \cline{1-2}
$E_{nei}$ & Num. of neighbor nodes \\ \cline{1-2}
$\mathcal{S}_e^{(c)}$ & \multicolumn{1}{l}{\begin{tabular}[c]{@{}l@{}}State space of the frontier node $e$ using \\  comm. resource $c$ \end{tabular}}   \\ \cline{1-2}
$s^{(c)}_{\eta,i}$ & \multicolumn{1}{l}{\begin{tabular}[c]{@{}l@{}}$i$-th element of the state space for\\   neighbor node $\eta$ \end{tabular}}   \\ \cline{1-2}
$Q(s_t,a_t)$ &  \multicolumn{1}{l}{\begin{tabular}[c]{@{}l@{}}Q-value corresponds to state $s_t$ and action $a_t$\end{tabular}} \\ \cline{1-2}

\end{tabular}}
\end{adjustbox}
\end{table}

\section{Deep Reinforcement Learning Agent for Multi-Flow Routing} \label{sec:DRL} 
In this work, we propose a DRL-based approach to solve the optimization problem (\ref{eq:objective function}) by exploiting a single DRL agent that is shared by all nodes. Specifically, we consider a MARL where a single, pre-trained DRL agent is deployed across all nodes, as opposed to different DRL agents being trained at different nodes, to significantly decrease the number of parameters and time to train. To further improve the scalability and generalizability, the DRL agent is trained with diverse flows and network topologies to ensure the DRL agent is robust to different scales of networks, numbers of active nodes, and mobility behaviors.

We train our DRL agent to work across all communication technologies and associated subbands. A naive way to do this is to train the DRL agent by aggregating features from all communication resources and neighbor nodes, and deciding the optimal next node and communication resource. However, this approach significantly increases the state and action spaces for the DRL agent resulting in slower convergence during training. Therefore, in this paper, we use a DRL agent for one communication resource at a time, using information only from neighbor nodes to estimate Q-values for each communication resource and neighbor node pair. Then, the agent selects the best node and corresponding communication resource with the highest estimated Q-value. This allows the agent to have a smaller state space and fewer actions to consider at each decision point, which helps the agent to converge faster to the near-optimal decision.
Recall that the same DRL agent is used not only for all nodes but also for all communication resources.

Our proposed DRL-based routing method consists of two parts, as seen in Fig. \ref{fig:framework}: 1) neighbor node set selection, 2) route decision and resource allocation. First, the neighbor node set selection process determines the neighbor nodes that are being considered as candidates at each relay node during the route decision process. In \cite{cui2021scalable}, a distance-based selection policy was considered in a distance-based channel model topology using a single communication technology. However, in this work, each node is equipped with multiple communication technologies and corresponding subbands, each with different channel characteristics; thus, there is no straightforward policy that selects the best subset of neighbor nodes as relay nodes.
Therefore, we consider three different approaches to select neighbor nodes, such as distance-based, channel-based, and rate-based approaches. The aim of this process is to select potential relay nodes to exchange information and reduce the communication overhead by limiting the information exchange only to the subset of relay nodes. Then, during the route decision and resource allocation process, the DRL agent uses the obtained information to estimate the Q-values and decide the node to transmit to and the communication resource to use.

\subsection{Neighbor Node Set Selection Approaches}\label{sec:node selection}
Since our proposed routing approach is a decentralized approach where each node has only information about its own neighbor nodes, it is important to decide which nodes to obtain information from. Also, due to the communication overhead that occurs while exchanging information, carefully selecting a smaller number of neighbor nodes is crucial for communication and network performance since the communication overhead scales proportionally with the number of neighbor nodes. To this end, we consider three different strategies, including distance-based, channel-based, and rate-based approaches, to select $E_{nei}$ neighbor nodes out of all $E$ number of relay nodes at the frontier node $e$ as potential candidate nodes for the route decision process where the selection of the appropriate number of $E$ depends on multiple factors such as node density of the topology and DRL algorithm. Note that the frontier node refers to the node that is currently making the decision.

\subsubsection{Distance-based neighbor node set selection}
Here, the frontier node $e$ selects the closest $E_{nei}$ neighbor nodes based on the distance regardless of the channel conditions. 

\subsubsection{Channel-based neighbor node set selection}
In this case, the frontier node $e$ computes the average channel gain across all communication resources with each neighbor node $k$ as $\frac{1}{|\mathcal{C}|}\sum_{i=1}^{M}\sum_{j=1}^{B^{(i)}}|h^{c_{(i,j)}}_{e,k}|$
. Then, the frontier node selects $E_{nei}$ nodes with the highest average channel gain as the neighbor nodes.

\subsubsection{Rate-based neighbor node set selection}
Here, the frontier node $e$ calculates the rate $\frac{1}{|\mathcal{C}|}\sum_{i=1}^{M}\sum_{j=1}^{B^{(i)}}R^{c_{(i,j)}}_{e,k}$ for node $k$. Then, it selects $E_{nei}$ nodes with the highest rate as the neighbor nodes. Note that this approach considers not only the channel effect but also interference from other hops and available bandwidth.

\subsection{Action Space}
Now, we define the action space for the DRL agent for each communication resource $c$ after neighbor nodes are selected. The set of all possible actions that can be made at the frontier node $e$ is given by
\begin{align} 
	\mathcal{A}_{e}^{(c)} = \{ a_{\eta_1}^{(c)}, a_{\eta_2}^{(c)},\cdots, a_{\eta_{E_{nei}}}^{(c)}   \},
\end{align} 
where $a_{\eta_i}^{(c)}$ is the action that frontier node $e$ transmits data to $i$-th neighbor node $\eta_i$ using communication resource $c$ and $E_{nei}$ is the number of neighbor nodes. Note that how to select neighbor nodes $(\eta_1,\cdots, \eta_{E_{nei}})$ significantly impacts the performance of the DRL agent, as it changes both the size and composition of the action space.

\subsection{State Space}

\begin{figure}[t]
    \centering  \includegraphics[scale=0.20]{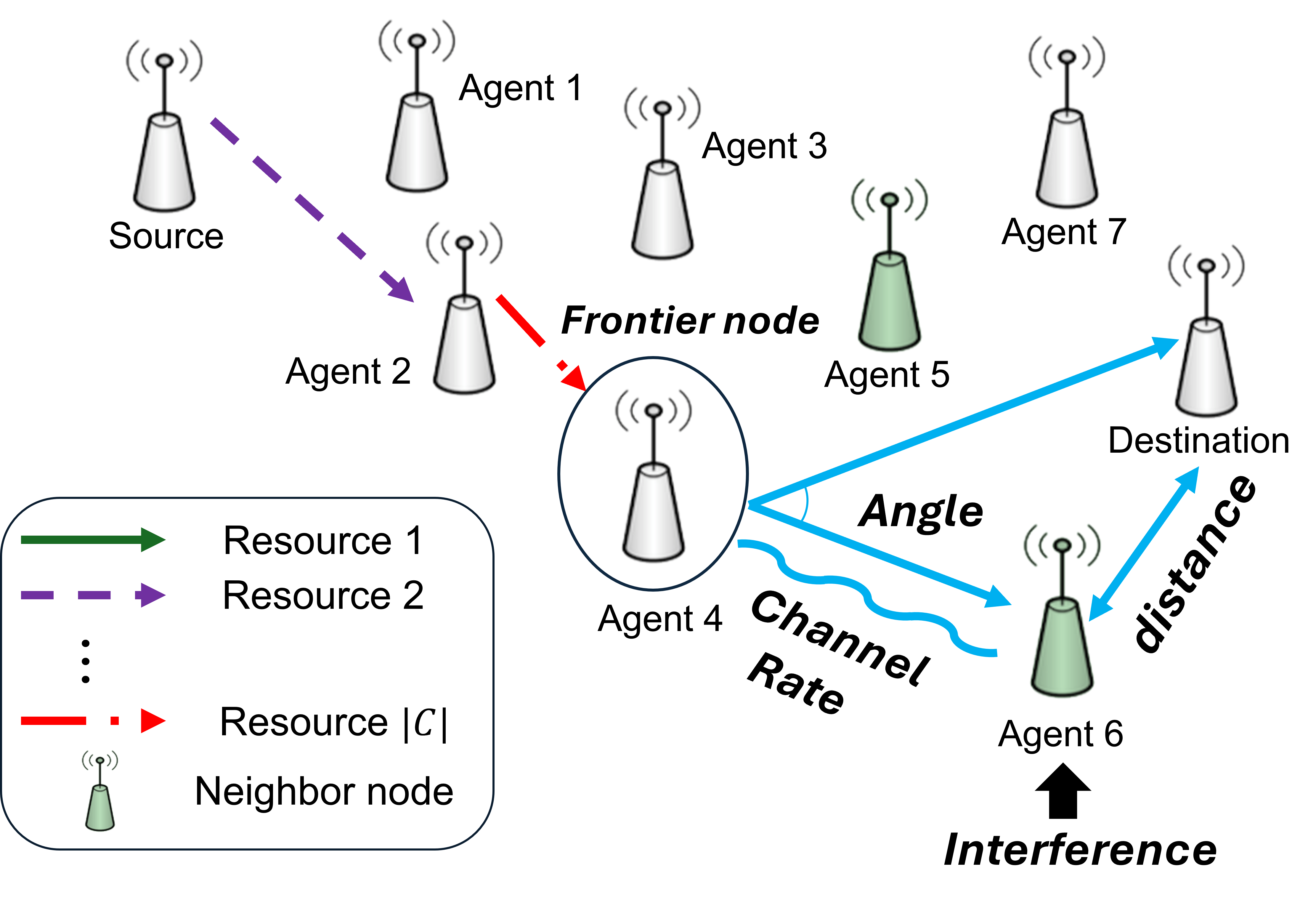}
    \caption{States that DRL agent at the frontier node gathers such as distance, angle, channel gain, interference, and rate.}
    \label{fig:states}
\end{figure}

The state space $\mathcal{S}^{(c)}_{e}$ of the frontier node $e$, corresponding to a communication resource $c$, needs to capture the representative features of neighbor nodes for effective action selection. Since the goal of our DRL agent is to maximize the rate of the route, we consider the following as the elements of the state space for the neighbor node $\eta$, as presented in Fig. \ref{fig:states}.
\begin{itemize}
     \item $s^{(c)}_{\eta,1}$: Distance between the neighbor node $\eta$ and the destination.
     \item $s^{(c)}_{\eta,2}$: Angle difference between the frontier node $e$ to the destination and to the neighbor node $\eta$. 
    \item $s^{(c)}_{\eta,3}$: Channel gain between the neighbor node $\eta$ and the frontier node $e$.
    \item $s^{(c)}_{\eta,4}$: Total interference the neighbor node $\eta$ is experiencing for communication resource $c$.
    \item $s^{(c)}_{\eta,5}$: rate between the neighbor node $\eta$ and the frontier node $e$
\end{itemize}
The state representation incorporates both geometric and wireless link information of neighbor nodes. Specifically, the geometric features, denoted as $s^{(c)}_{\eta,1}$ and $s^{(c)}_{\eta,2}$, capture the relative proximity and direction of a neighbor node with respect to the destination. In addition, the wireless link features, $s^{(c)}_{\eta,3}$, $s^{(c)}_{\eta,4}$, and $s^{(c)}_{\eta,5}$, characterize the quality of the communication link between the frontier node and the neighbor node. By defining this five-dimensional feature space for each neighboring node, the overall state space used for action selection at the frontier node with respect to the communication resource $c$ consists of $5\cdot E_{nei}$ factors, expressed as
\begin{align} \label{eq:state}
	\mathcal{S}_{e}^{(c)} = \{ s^{(c)}_{\eta_1,1},s^{(c)}_{\eta_1,2},\cdots, s_{\eta_{E_{nei}},5}^{(c)}   \}.
\end{align}



It is important to note that information such as coordinates, cumulative interference, and channel gain can be easily shared among nodes through the control and non-payload communications (CNPC) link, as outlined in \cite{cui2021scalable, zhang2024decentralized}. These types of data are commonly utilized in routing and resource allocation tasks where the information exchange is generally confined to neighboring nodes. Therefore, the DRL agent introduces minimal additional overhead relative to existing algorithms when gathering local feature data.
\begin{figure*}[t]
    \centering  \includegraphics[scale=0.20]{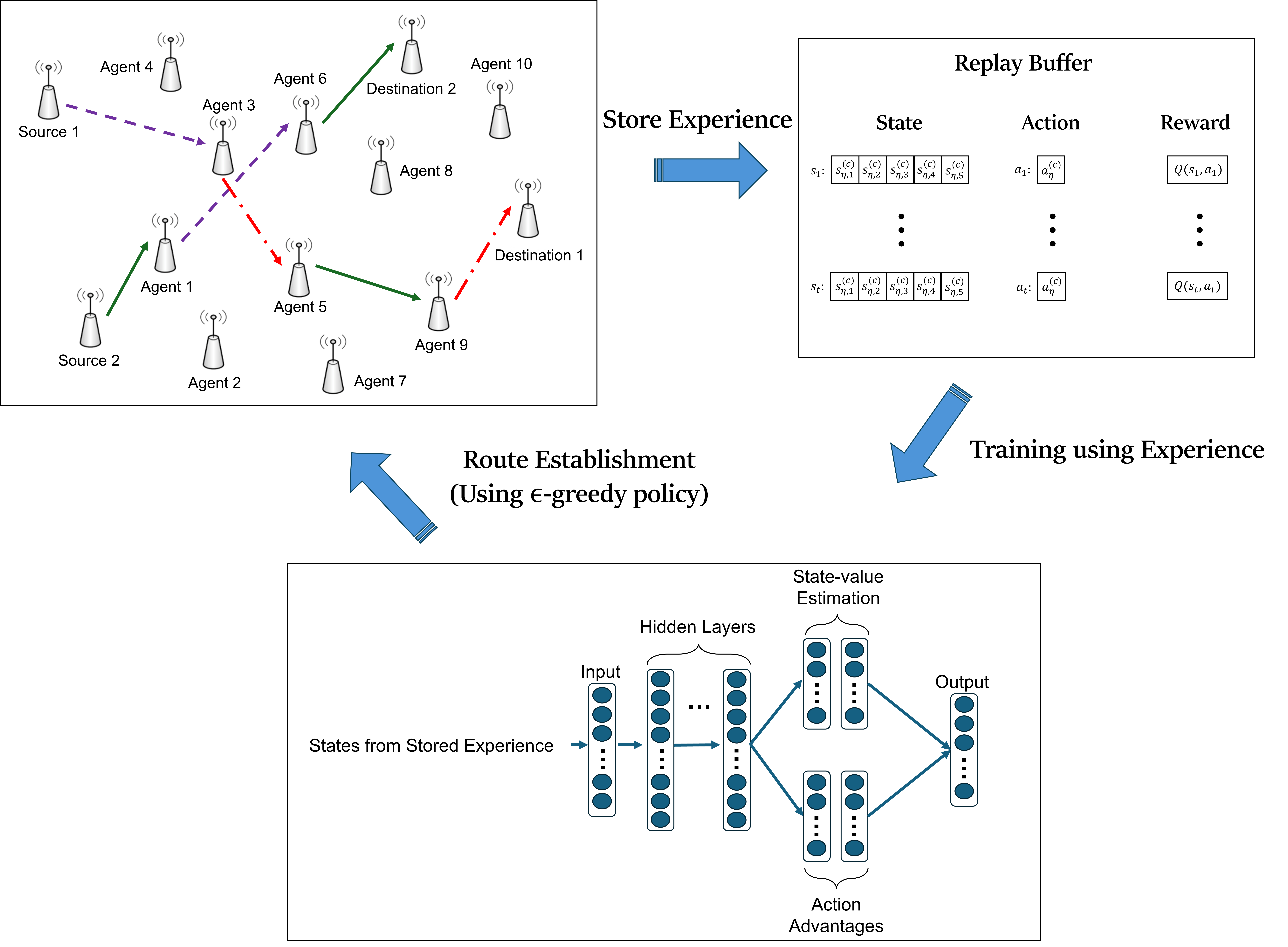}
    \caption{Training process of DDQN with experience replay. The route is established using $\epsilon$-greedy policy. After the route is established, (state, action, reward) tuple is saved and used during training. }
    \label{fig:drl}
\end{figure*}

\subsection{Reward function}
In this paper, we implement DQN \cite{mnih2015human} to approximate the Q-value function by exploiting DNNs. The Q-value, estimating the value of a state-action pair, is used to decide the action by the agent, which selects the action associated with the highest Q-value. Conventionally, the Q-value $Q(s_t, a_t)$ is updated during the training using the immediate reward and the future reward as follows:
\begin{align}
    Q(s_t, a_t) = Q(s_t, a_t) + \alpha [ r_t + \gamma \max_{a'} &Q(s_{t+1}, a') \nonumber\\
    &- Q(s_t, a_t) ],
\end{align}
where $r_t$ is the immediate reward, $Q(s_{t+1}, a')$ is the future reward from the next state $s_{t+1}$, $\alpha$ is the learning rate, $\gamma$ is the discount factor, and $(s_t,a_t)$ is the action $a_t$ state $s_t$ pair at time $t$.

To compute the Q-value, we first need to develop a reward function that is tailored to our goal, which is to maximize the sum rate of all routes. However, due to the decentralized and sequential nature of the decision-making process of our algorithm, it is hard to incorporate information about rates from other flows into the reward function. Thus, we design our DRL agent to maximize the end-to-end rate of its own route considering interferences from other flows. To this end, we set the reward function as the end-to-end rate of the route, determined by the bottleneck rate of the established route. Nevertheless, this choice of reward introduces a challenge that hinders the agent from finding the optimal solution since the earlier hops can dominate the bottleneck rate. Specifically, the immediate and the future rewards in the Q-value function are set as the rate from the earlier hops, making the action chosen at the current step irrelevant.

To circumvent this issue, we define the Q-value with only the future reward which is defined as the bottleneck rate after the route has been established, similar to the approach done in \cite{cui2021scalable}.  Specifically, during training, the agent observes the state $s_t$ and the action $a_t$ until it reaches the destination. Then, after the route is fully established, state-action pairs along the route are stored with the future reward, which is the bottleneck rate of the established route $G^{(f)}$, as follows. 
\begin{equation}\label{eq:Q-value}
    Q(s_t, a_t) = \underset{ e^{(f)}_i \in G^{(f)}}{\min} R^{c^{(f)}_{i}}_{e^{(f)}_{i},e^{(f)}_{i+1}}.
\end{equation}

\subsection{Architecture of the DRL agent} \label{sec: architecture DRL}
Given that the state features are represented as real-valued variables and that wireless networks exhibit inherently dynamic behavior, the DRL agent must be able to generalize across previously unseen environments with distinct state characteristics. To address this, we adopt a DQN framework, where DNNs are employed to estimate the Q-values corresponding to state–action pairs. We leverage the dueling DQN (DDQN) architecture \cite{wang2016dueling}, in which the network is decomposed into two separate streams: one estimating the state value function and the other capturing the action advantage function, as illustrated in Fig. \ref{fig:drl}. The key idea of DDQN is to learn how good the state is and which actions are better than others, then merge them into Q-values. During training, the state input is represented as a vector of length $5 E_{nei}$ as in Equation (\ref{eq:state}), while the associated Q-values for each action are computed according to Equation (\ref{eq:Q-value}). 

During DDQN training, we employ the experience replay mechanism introduced in \cite{mnih2015human}, where experiences are collected using an $\epsilon$-greedy policy \cite{sutton2018reinforcement}. In this policy, each node selects an action based on the Q-value predicted by the DDQN (exploitation) with probability $1-\epsilon$, while with probability $\epsilon$ it selects a random action from the action space (exploration) to encourage the discovery of unvisited actions. The $\epsilon$-greedy policy thus yields diverse state–action pairs, for which the corresponding bottleneck rate is computed once the route is fully established. These experiences are stored as $(\text{state}, \text{action}, \text{reward})$ tuples in the replay buffer, which are subsequently sampled for training. To ensure convergence, we linearly decrease $\epsilon$ and ultimately fix $\epsilon=0$ at the end of training, allowing the agent to operate without further exploration. Specifically, we set $\epsilon_t = 1 - t/T$ where $T$ denotes the total number of training episodes for $\epsilon$-greedy method and $t$ denotes the current number of training episode. Once training is complete, the DDQN is applied to each communication resource, and the frontier node selects the action corresponding to the maximum estimated Q-value across all communication resources and neighbor nodes.

\section{Simulation Results} \label{sec:simulations}

\subsection{Simulation Environment} \label{sec:simulation environment}
During simulations, we train and evaluate our DRL agent using datasets created based on two distinct environments: 1) a small scale high-fidelity ray tracing-based simulation of a cluttered propagation environment (see Fig. \ref{fig:simulation}), and 2) a larger scale path loss model-based simulation. The high fidelity ray-tracing simulation dataset is employed to capture realistic scenarios, incorporating detailed channel effects across different communication technologies. On the other hand, the path loss model-based environment provides a simplified, distance-dependent abstraction that is highly flexible. For example, it allows for easier adjustment of network size and node density, as well as more straightforward tracking of mobile nodes. Note that both datasets account for the unique channel characteristics of different communication technologies with different levels of fidelity.

\subsubsection{Ray tracing-based environment} 
\begin{figure}[t]
    \centering  \includegraphics[scale=0.3]{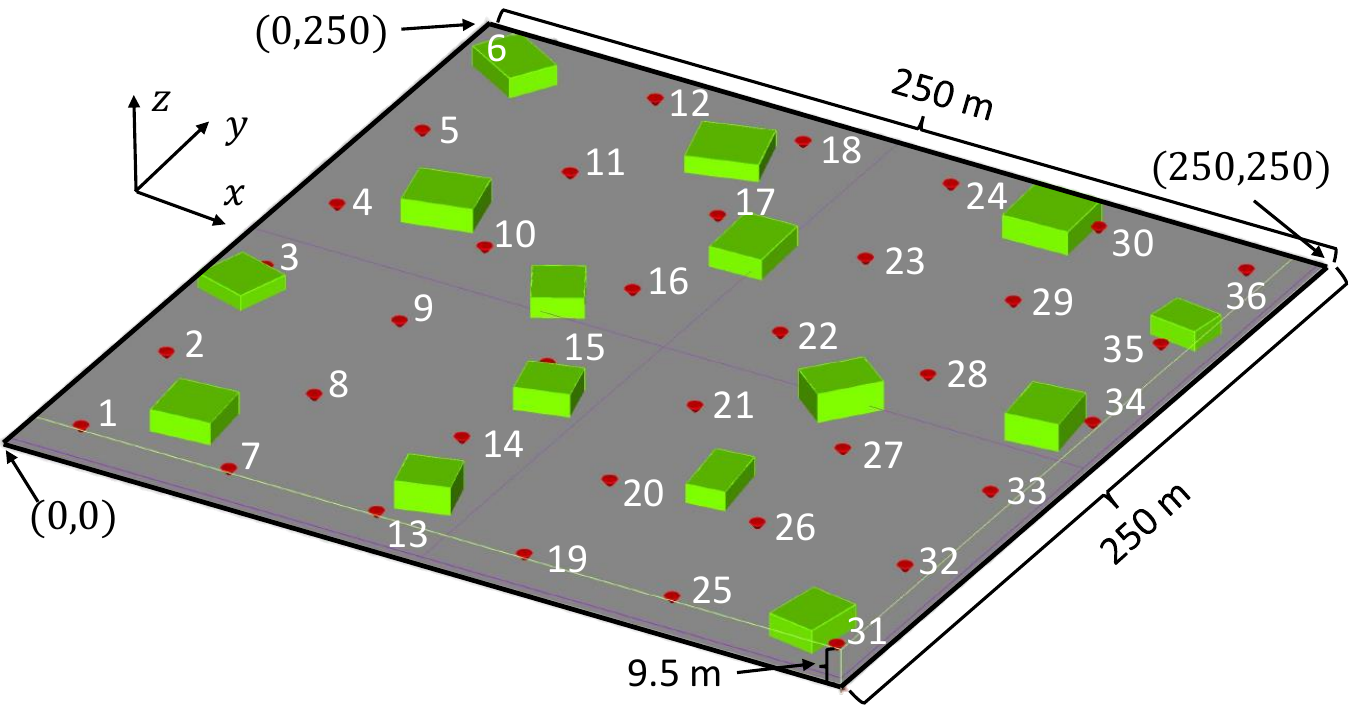}
    \caption{3D ray tracing-based simulation environment with 36 nodes.}
    \label{fig:simulation}
\end{figure}
We consider a three-dimensional multi-building simulation environment of size $(250 \times 250 \times 9.5)$ m$^3$, consisting of concrete buildings (green boxes) and 36 legitimate nodes (red dots), as illustrated in Fig.~\ref{fig:simulation}. Each legitimate node is equipped with three wireless communication technologies, denoted by $m_1$, $m_2$, and $m_3$, operating at center frequencies $\tau_{m_1} = 400$~MHz, $\tau_{m_2} = 900$~MHz, and $\tau_{m_3} = 2.4$~GHz, respectively. The bandwidth assigned to a subband of technology $m_i$ is given by $\Omega^{(m_i)} = 0.01 \tau_{m_i} / B^{(m_i)}$~MHz, where $B^{(m_i)}$ represents the number of subbands. In this setup, the total bandwidth of each communication technology is defined as 1\% of its center frequency. Different values of $B^{(m_i)}$ are considered during simulations to capture diverse scenarios. Channel data for all three communication technologies and all 36 nodes are computed over a grid of densely sampled receiver locations using a high-fidelity ray-tracing-based approach \cite{EMCUBE}. To generate diverse network topologies during training and testing, we randomly select 25 nodes from the set of all legitimate nodes excluding source-destination pairs.  

\subsubsection{Path loss model-based environment}
We consider a simulation environment consisting of different ranges of communication such as very high frequency (VHF) and ultra high frequency (UHF), operating at 30-300MHz and 300MHz-3GHz, respectively. Specifically, for our simulations, we assume that each node is equipped with seven wireless communication technologies, operating at center frequencies $(\tau_{m_1},\tau_{m_2},\tau_{m_3},\tau_{m_4},\tau_{m_5},\tau_{m_6},\tau_{m_7}) = (40, 80, 200, 400, 800, 2000, 3000)$ MHz where we implement channel models for $\tau_{m_1},\tau_{m_2}$ from \cite{andrusenko2008vhf} and $\tau_{m_3},\tau_{m_4},\tau_{m_5},\tau_{m_6}$ from \cite{perez2002path}, and $\tau_{m_7}$ from \cite{bjornson2019intelligent}. Similar to the ray tracing-based environment, the bandwidth assigned to a subband of technology $m_i$ is given by $\Omega^{(m_i)} = 0.01 \tau_{m_i} / B^{(m_i)}$~MHz and the total bandwidth of each communication technology is defined as 1\% of its center frequency. During our simulations, we consider different densities of the network topology where we consider the size of (2$\times$2) km$^2$, and 27 and 90 relay nodes in the topology to represent a sparse and dense network, respectively. During the training phase, nodes are randomly placed to make the DRL agent robust to the topology randomness. 

\subsubsection{DDQN specification and training} 
We consider DDQN architecture consists of input layer of size $5E_{nei}$, 3 hidden layers with 300 nodes, two hidden layers with 300 and 150 nodes for each layer for the state-value estimation part, two hidden layers with 300 and 150 nodes for each layer for the action advantages part, and the output layer of size $E_{nei}$. 
Here, the input and output dimensions solely depend on the number of neighbor nodes $E_{nei}$. 

To train DDQN with multiple flows, we store experience as explained in Section \ref{sec: architecture DRL} where the flows except the last flow are established using one of the benchmark schemes, closest to destination scheme, which is explained in the following section. Then, the last flow is established using DDQN and stores the state-action pair as an experience. Since the existing flows affect the experience stored to train DDQN, it is important to investigate how the difference in the number of flows during the training and test phases affects DDQN performance.

\begin{algorithm}[t]

\DontPrintSemicolon
\SetAlgoNoEnd
\SetAlgoLined
Define source $S$ and destination $D$ nodes.\\
Define vertices (nodes) $V$ for a given network topology. \\
Define frontier node $F = S$.\\
\While{$F \ne D$}{
Create a graph $(V,W)$ where edges $W$ are defined as $w_{i,j} = \max_{c} R^{c}_{i,j}$ between node $i$ and $j$.\\
Obtain a route $[F, N_1, N_2, \cdots, D]$ from the widest path algorithm.\\
Update $F = N_1$.
}
Route = All intermediate frontier nodes $F$.
\caption{Widest path-based algorithm \label{alg:widest path}} 
 \end{algorithm}

\subsection{Benchmark Schemes for Routing}
Since the optimal solution for multi-hop wireless routing with multiple flows under interference is infeasible, we evaluate our proposed DRL agent against several benchmark schemes as follows.
\begin{itemize}
    \item \textit{Best direction to destination}: Select the neighbor $\eta$ with the best direction such that the state $s^{(c)}_{\eta,2}$ is smallest.
    \item \textit{Closest to destination}: Select the neighbor $\eta$ closest to the destination node such that $s^{(c)}_{\eta,1}$ is smallest. 
    \item \textit{Least interfered}: Select the neighbor and communication resource with the least interference such that $s^{(c)}_{\eta,4}$ is smallest.
    \item \textit{Largest data rate}: Select the neighbor and communication resource with the highest link capacity such that $s^{(c)}_{\eta,5}$ is highest.
    \item \textit{Destination directly}: Transmit to the destination directly from the source.
    \item \textit{Widest path-based algorithm}: Obtain a route from the frontier node to the destination using the widest path algorithm where the graph is defined with edge weights corresponding to the maximum rate of each link across all communication resources. The frontier node then selects only the next hop and communication resource from this route. The graph is subsequently updated to account for interference from the already selected transmitting nodes. The detailed procedure is provided in Algorithm \ref{alg:widest path}.

\end{itemize}
Note that for \textit{Best direction to destination}, \textit{Closest to destination}, and \textit{Destination directly} schemes, the communication resource is selected with the highest rate between the frontier node and the selected neighbor node.




\subsection{Multi-Flow Routing Performance} 

We first investigate the performance of the proposed DRL agent for HWNs with multiple flows and compare it with other benchmark schemes for both path loss model-based and ray tracing-based environments. To represent the average performance of each approach, the results are evaluated with 1000 different topologies based on the sum rate of multiple flows and presented with a cumulative distribution function (CDF). Furthermore, to promote fairness among multiple flows since routes established earlier are more susceptible to interference from subsequently established flows, we implement a route re-establishment procedure after the initial routing. Specifically, routes are re-established in descending order of their achieved rates, and this process is iteratively repeated four times throughout the simulation to enhance overall fairness and performance balance among the flows.

\begin{figure}[t]
    \centering  \includegraphics[scale=0.4]{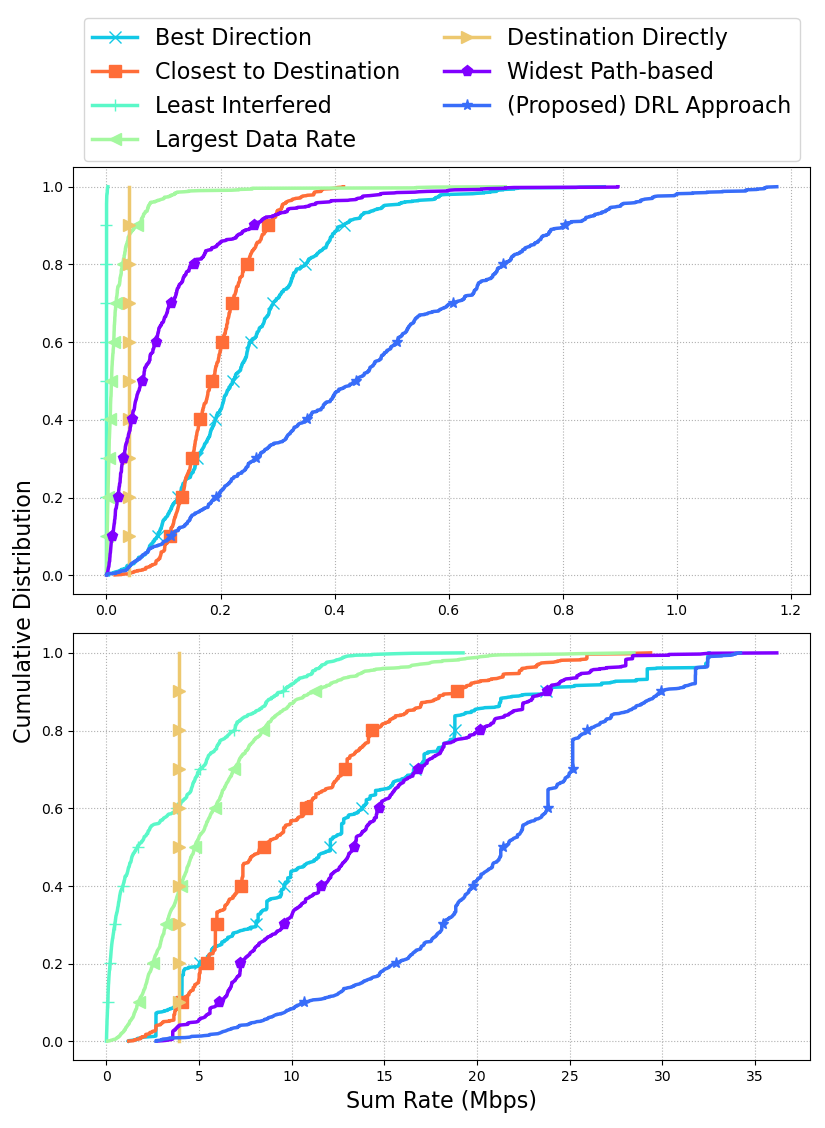}
    \caption{Performance of DRL agent with 2 flows compared to other benchmark schemes using path loss model-based environment (top) and ray tracing-based environment (bottom).}
    \label{fig:2flow_benchmarks}
\end{figure}

In Fig. \ref{fig:2flow_benchmarks}, we consider $F=2$ concurrent flows for both the path loss model-based environment and ray tracing-based environment where the transmit power of each node is set as $P = 0$~dBm and noise power $N_0 = -110$~dBm. For the path loss model-based environment, we consider $E = 90$ relay nodes inside the (2$\times$2)~km$^2$ size topology and set the size of neighbor nodes as $E_{nei} = 30$ and $B^{(m_i)} =1$ for all seven communication technologies. As a result, it is seen in the top figure of Fig. \ref{fig:2flow_benchmarks} that our proposed DRL approach significantly outperforms other benchmark schemes. A similar result has been shown in the bottom figure where the ray tracing-based simulation dataset is considered with $E_{nei} = 5$, $B^{(m_i)} =5$ for all three communication technologies, and $E = 25$. 

\begin{figure}[t]
    \centering  \includegraphics[scale=0.48]{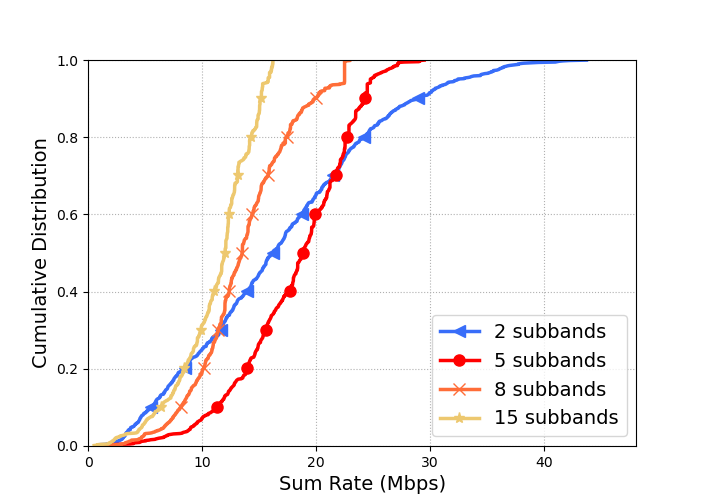}
    \caption{Analyzing the performance with different numbers of subbands in the ray tracing-based environment.}
    \label{fig:EMCUBE_2flows_subbands}
\end{figure}

To evaluate the impact of the number of subbands for each communication technology on the sum rate, we vary the number of subbands for the ray tracing-based environment, as shown in Fig. \ref{fig:EMCUBE_2flows_subbands}. The case with $B^{(m_i)}=5$ subbands outperforms the case with fewer $B^{(m_i)}=2$ subbands on average, yielding 18.2 Mbps and 16.2 Mbps, respectively. This result indicates that mitigating interference by increasing the subbands surpasses the decrease in the bandwidth allocated for each subband. 
Furthermore, the $B^{(m_i)}=5$ subbands case outperforms the cases with $B^{(m_i)}=8$ and $15$ subbands, showing that the decrease in the bandwidth allocated for each subband dominates, thus reducing the sum rate. This observation shows a trade-off: increasing the number of subbands reduces interference, thereby enhancing the rate, while decreasing the number of subbands increases the bandwidth allocated per subband, which also contributes to a higher rate. The optimal performance is achieved when these opposing effects are balanced. 


\begin{figure}[t]
    \centering  \includegraphics[scale=0.43]{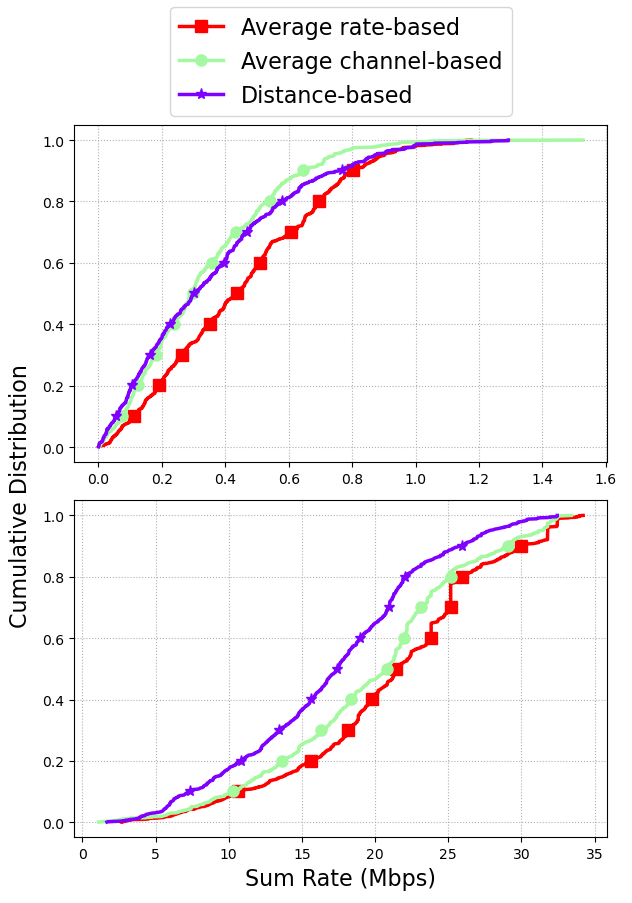}
    \caption{Different neighbor node set selection methods with 2 flows for path loss model-based environment (top) and ray tracing-based environment (bottom).}
    \label{fig:neighbor selection}
\end{figure}



Different approaches of selecting neighbor nodes, as discussed in Section \ref{sec:node selection}, are investigated in Fig. \ref{fig:neighbor selection}. Here, analyses are based on the same scenarios that are used for Fig.~\ref{fig:2flow_benchmarks} and the DRL approach is considered. The results show that the average rate-based approach performs the best. This is because this approach not only considers the interference effect that the neighbor node is experiencing, but also the channel effect and the allocated bandwidth. Further, it is observed in the top figure that the average channel-based and distance-based yields similar results because the channel model that is considered during this simulation is a path loss model. In such a model, the distance-based and channel-based neighbor node sets become the same, differing only slightly due to the randomness introduced during the training process of the DRL agent. A comparable trend is observed in the bottom figure, where the distance-based approach underperforms relative to the other approaches due to the more realistic scenario considered in the ray tracing-based simulation.

\subsection{Routing Performance in the Presence of Node Mobility}
To evaluate the robustness of the DRL agent against node mobility, we test our approach in a network topology where a subset of relay nodes is mobile. Specifically, we consider a path loss model-based environment with 27 relay nodes, of which 20 are moving. During the simulations, we employ a pre-trained DRL agent that was trained under static conditions (i.e., without any node mobility) and evaluate its performance in the presence of node mobility. To present the average performance, we evaluate over 500 different topologies and compute the mean performance across all simulations.

\begin{figure}[t]
    \centering  \includegraphics[scale=0.35]{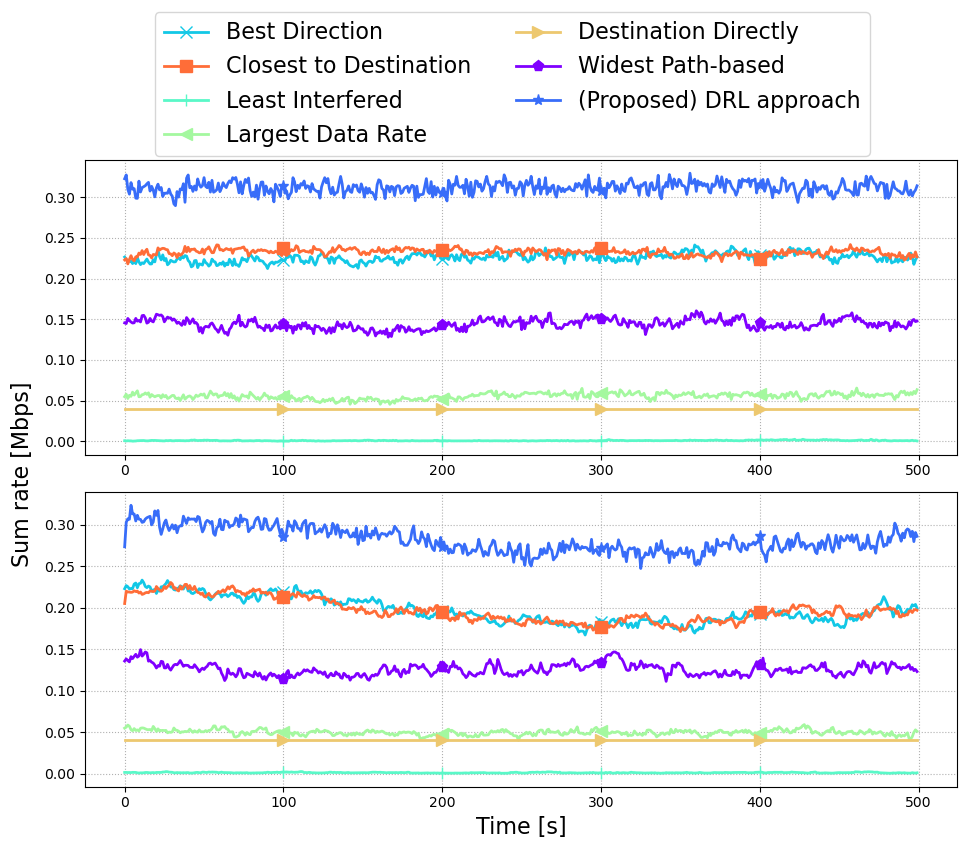}
    \caption{DRL agent robustness to node mobility (top) random walk model (bottom) random waypoint model.}
    \label{fig:node mobility}
\end{figure}
We consider two mobility models during simulation, the random walk mobility model \cite{einstein1956investigations} and the random waypoint mobility model \cite{broch1998performance}. In the random walk mobility model, each mobile node randomly selects a direction (between [0, $2 \pi$)) and a speed (between [0, 5] m/s) every second. Once the node reaches the boundary it reflects off the boundary and continues moving in the new reflected direction. For the random waypoint mobility model, each mobile node randomly selects a destination within the topology and moves toward it at a speed randomly chosen from the range [0, 5] m/s, with the speed reselected every second. Note that, once a mobile node reaches its destination, a new destination is selected. 

The robustness of the DRL approach under the random walk and waypoint mobility models is illustrated in Fig. \ref{fig:node mobility}, where the performance is evaluated at one-second intervals. Note that each node independently makes decisions based on updated information gathered every second.
The results show that, despite node mobility, the DRL approach consistently outperforms benchmark schemes and maintains stable performance for both mobility models. However, there exists a slight decrease in performance for our DRL approach over time for the random waypoint mobility model compared to the random walk mobility model.

\begin{figure}[t]
    \centering  \includegraphics[scale=0.41]{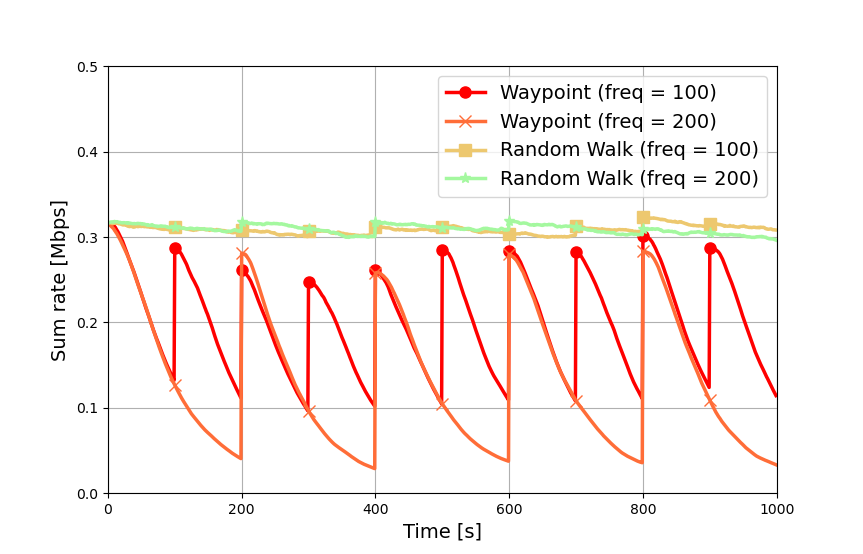}
    \caption{Longevity of the solution from the DRL agent.}
    \label{fig:DIST_2flows_stale}
\end{figure}

In contrast to making decisions at each node every second based on the information available at that moment, the impact of decision longevity is studied in Fig. \ref{fig:DIST_2flows_stale}, where each node updates its decision at a fixed frequency, i.e., every 100$s$ or 200$s$. It is observed that the performance of the DRL approach decreases almost linearly as time passes for the random waypoint mobility model and instantly goes up when the route is updated, while it remains more robust under the random walk mobility model. Therefore, it is crucial to estimate the mobility model in advance to determine the optimal frequency at which information should be obtained from neighbor nodes to re-establish the route.

\subsection{Generalizability of the DRL agent}



\begin{figure}[t]
    \centering  \includegraphics[scale=0.45]{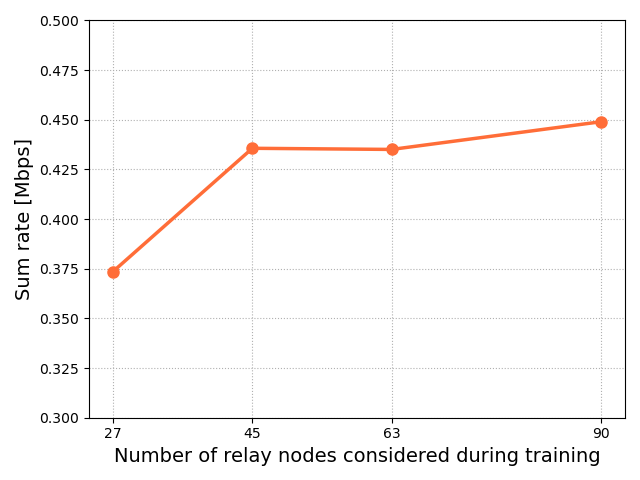}
    \caption{Performance of the DRL approach trained with a network with different numbers of relay nodes when tested in a dense network (90 relay nodes). }
    \label{fig:sparse dense}
\end{figure}

\begin{figure}[t]
    \centering  \includegraphics[scale=0.46]{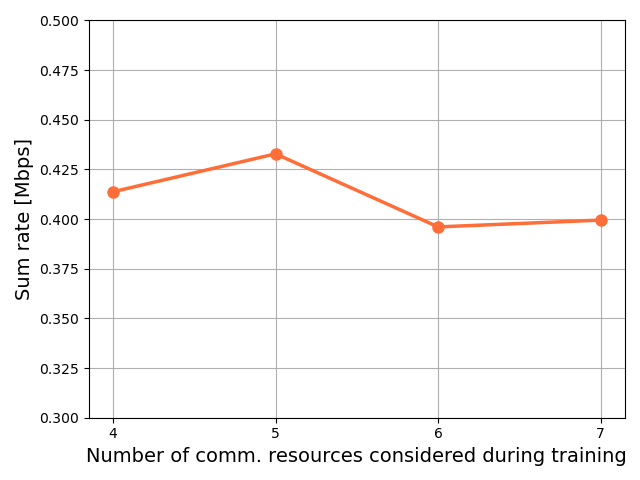}
    \caption{Performance of the DRL approach trained with a network with different numbers of communication resources equipped at each node when tested with the 7 communication resources equipped.}
    \label{fig:diff communication resources}
\end{figure}

To evaluate the generalizability of the proposed DRL approach to network variations during the testing phase, we assess the performance of a pre-trained DRL agent in a network configuration different from the one used during training. In Fig. \ref{fig:sparse dense}, we examine how changes in the number of relay nodes during the test phase influence the performance of the pre-trained DRL agent. In the simulation, two data flows are considered, and the sum rate is obtained from the configuration of $E_{nei} = 5, 10, 15, 20$ that yields the highest performance. The DRL agent is tested in a path loss model-based environment consisting of 90 relay nodes, while it is trained with networks containing the number of relay nodes indicated on the x-axis. The results show that the DRL agent trained with $E = 45$ and $E = 63$ relay nodes experiences only about a 3\% reduction in sum rate compared to the agent trained with $E = 90$ relay nodes. In contrast, the DRL agent trained with $E = 27$ relay nodes presents degraded performance, around 17\% reduction, as it was exposed only to sparse network conditions during training.

\begin{figure}[t]
    \centering  \includegraphics[scale=0.46]{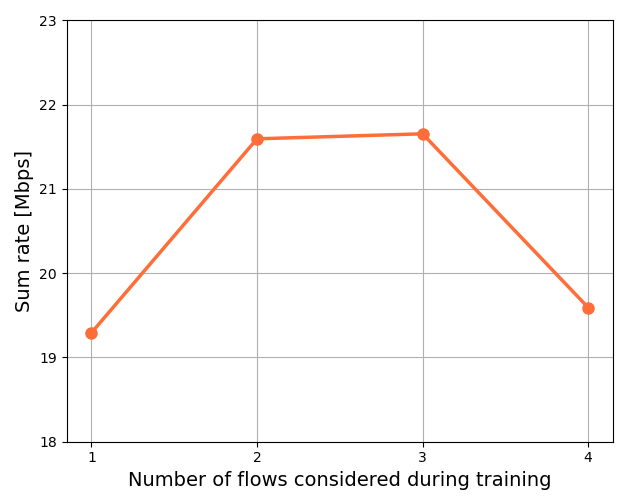}
    \caption{Performance of the DRL approach trained with a network with different numbers of flows in the network when tested with the 4 flows in the network.}
    \label{fig:diff flows}
\end{figure}

Similarly, the performance of a pre-trained DRL agent under varying numbers of communication resources is evaluated in Fig. \ref{fig:diff communication resources}. We consider the path loss model-based environment with $E = 60$ relay nodes with two data flows where each node selects $E_{nei} = 10$ neighbor nodes. Here, each node is equipped with $M$ communication technologies, each utilizing one subband (corresponding to the x-axis of the figure), with center frequencies selected from $(40, 80, 200, 400, 800, 2000, 3000)$ MHz in ascending order. Note that the number of communication resources equals the number of communication technologies since only one subband per technology is considered in this simulation. As shown in Fig. \ref{fig:diff communication resources}, the pre-trained DRL agent demonstrates robust performance even when evaluated under different numbers of communication resources, comparable to the case where both training and testing are performed with seven communication resources. This suggests that the proposed DRL approach is well generalizable to varying numbers of communication resources, as its input/output dimension remains unaffected by the number of communication resources, enabling it to effectively learn and adapt to the characteristics of different communication technologies.

The performance of the pre-trained DRL agent under different numbers of data flows is illustrated in Fig. \ref{fig:diff flows}. The evaluation is conducted within the same ray tracing-based environment used in Fig. \ref{fig:2flow_benchmarks}. As shown in Fig. \ref{fig:diff flows}, the pre-trained DRL agent maintains robust performance even when the number of flows during the testing phase differs from the training phase, which was based on four flows. Interestingly, a slight performance improvement is observed in the cases with two and three flows compared to the four flows case. This improvement can be attributed to the characteristics of the stored training data. As discussed in Section \ref{sec:simulation environment}, the stored data correspond only to the last flow during training, whereas, during testing, the DRL agent is applied to all flows. Consequently, the earlier flows experience minor performance degradation due to the data imbalance between training and testing phases. Overall, this result indicate that the proposed DRL-based routing approach effectively generalizes across diverse flow scenarios, thereby validating its adaptability and robustness in handling dynamic multi-flow network conditions.


\section{Conclusion and Future Work} \label{sec:Conclusion}
In this work, we presented a DRL-based routing framework for HWNs that addresses the challenges of multi-hop communication across diverse technologies and dynamic environments. By leveraging DQNs, our method enables each node to make joint decisions on communication technology, subband, and next-hop relay while accounting for both channel conditions and interference not only from other flows but also from hops within the flow. The proposed framework not only overcomes the limitations of conventional distance-based routing but also demonstrates strong adaptability to variations in mobility, node density, and number of flows. Simulation results confirm that the approach enhances scalability and achieves a higher end-to-end sum rate compared to various benchmark schemes. Moreover, the DRL agent exhibits robustness to relay node mobility and generalizability to network variations during the test phase, with only about a 3\% sum rate degradation observed when the number of relay nodes was altered. These findings highlight the potential of DRL-driven routing strategies to support efficient and robust communication in increasingly complex wireless networks.


Despite providing important analysis towards the robustness and generalizability of the DRL approach for multi-flow routing in HWNs, this paper still leaves various research directions open for future work. One of the important directions is to investigate the overall overhead, especially the communication overhead, that is needed during the training phase of the DRL approach to collect the dataset. Future work will also focus on fine-tuning the DRL framework to improve generalizability and performance by exploring different DRL approaches and transfer learning strategies to adapt to dynamic network conditions. Additionally, integrating domain knowledge such as channel state information and topology prediction into the learning process could further enhance decision-making efficiency.




\bibliographystyle{ieeetr}
\bibliography{references}

\end{document}